\documentclass[
reprint, 
 amsmath,amssymb,
prl,
nobalancelastpage
]{revtex4-2}

\usepackage{graphicx}
\usepackage{dcolumn}
\usepackage{bm}
\usepackage[utf8]{inputenc}
\usepackage[final]{changes}  

\begin{document}

\preprint{APS/123-QED}

\title{{Available Energy and Ground States of Convective Hydrodynamic and Hydromagnetic Instabilities} }%

\author{Kaixuan Fan}

\author{Yao Zhou}%
\email{yao.zhou@sjtu.edu.cn}
\affiliation{%
School of Physics and Astronomy, Institute of Natural Sciences, and MOE-LSC, Shanghai Jiao Tong University, Shanghai 200240, China
}%

\date{\today}

\begin{abstract}

We propose a method for predicting the nonlinear extent of convective instabilities in neutral and magnetized fluids. The method extends Gardner's restacking algorithm, which computes the available energy and ground states of collisionless plasmas in phase space, and combines it with Lagrangian relaxation, where fluid elements find lower-energy equilibria while preserving local invariants. 
For the incompressible Rayleigh--Taylor instability, the problem is equivalent to Gardner's formulation and the restacking algorithm directly applies in configuration space. 
To treat compressibility in a general setting, we follow restacking with Lagrangian relaxation to obtain the ground state, and the results show excellent agreement with direct numerical simulations.
Successful extension to the $m=0$ interchange instability in a Z-pinch demonstrates the method's potential as a general framework for estimating the nonlinear extent of convective instabilities, which can facilitate the design and operation of fusion reactors.

\end{abstract}

\maketitle

\textit{Introduction}---Convective instabilities are ubiquitous in fluids and plasmas, arising whenever an unfavorable stratification in an effective gravitational field allows elements to interchange and release potential energy \cite{Chandrasekhar1961}. 
In neutral fluids, a canonical example is the Rayleigh--Taylor instability (RTI), which drives rapid mixing and often dominates the nonlinear evolution in laboratory, geophysical, and astrophysical flows \cite{ZHOU20171}. In magnetized plasmas, the closely related interchange and ballooning instabilities play an important role in setting confinement limits and mediating transport \cite{freidberg2014ideal}.

Linear stability is well described by the energy principle \cite{Bernstein1958} and typically treated as a hard constraint on the design and operation of fusion reactors. 
Yet the nonlinear outcome of a linear instability can bifurcate between a major crash or benign saturation. 
In the tokamak core, the quasi-interchange instability can trigger sawteeth or maintain a helical steady state \cite{Wesson1986,Menard2006,Jardin2015,Jardin2020}.
In the periphery, the edge-localized modes (ELMs) can induce intense heat pulses to the plasma-facing components or manifest as small, ``grassy'' oscillations \cite{Zohm1996,Kirk2006,Xu2019}. 
In stellarators, the $\beta$-limits predicted by linear stability  theory tend to be soft, even though large-scale collapses still occur at times \cite{Weller2006,Ohdachi2017,Zhou2024}.
Despite significant theoretical efforts to characterize the nonlinear behaviors of these instabilities \cite{Wilson2004,Zhu2009,Xi2014,Ham2016,Chu_2026}, a general framework remains elusive. 
In particular, a practical model that can efficiently estimate the nonlinear extent of the instabilities without invoking expensive simulations can expand the design and operation space of fusion reactors by relaxing linear stability  constraints.

 {The available energy can serve as such a dynamics-agnostic nonlinear measure. It is defined as the excess energy above a “ground state”, obtained by minimizing the energy over rearrangements of discrete parcels constrained by the invariants of the system. The nature of these parcels and the invariants depend on the physical model. Gardner conceived this idea for kinetic instabilities in collisionless plasmas by restacking incompressible phase-space elements \cite{gardner1963bound}.} 
This algorithm has been formalized \cite{dodin2005variational,helander2017available} and successfully applied to microinstabilities and the resulting turbulent transport in magnetic-fusion plasmas \cite{helander2020available,mackenbach2022available,sun2024available,Mackenbach2025,MacKenbach2026}, and also motivated studies of phase-space engineering \cite{Kolmes2024,Qin2025}. 
\added{Related constructions have also appeared beyond plasma physics, including in quantum thermodynamics, where the available energy is known as ergotropy \cite{Campisi_2026,campisi2026ergotropicrearrangementphasespace}, and in galactic dynamics  \cite{Lemou2012,Wiechen1988}.}
In meteorology, Lorenz \cite{Lorenz1955}  {conceptualized and calculated the available (potential) energy of circulation in stratified air by similarly rearranging adiabatic fluid parcels. 
However, the treatment of compressibility in this line of work \cite{Lorenz1979,Hieronymus2015,Stansifer2017,Harris2018,Hosking2025} relies on reducing the force balance condition to a simple algebraic relation between the pressure and the ``mass coordinate" adopted (cf.\ End Matter).
This convenience is typically absent in many other systems, which motivates a general treatment of compressible fluids.} 
A closely related idea is Lagrangian relaxation, where fluid elements rearrange to find energy-minimizing equilibria while preserving local invariants including mass, entropy, and magnetic flux \cite{Newcomb1962,Craig1986,Candelaresi2014,Zhou2014}. Its solution is given in terms of a continuous configuration map, which is more restrictive than Gardner's discontinuous restacking operation and hence cannot capture nonlinear convection alone.

In this letter, we present a novel method for calculating the available energy and ground states of convective instabilities by extending Gardner’s restacking and combining it with Lagrangian relaxation,  {offering a general treatment of compressibility}. 
We first show that the incompressible RTI admits an equivalent restacking formulation in configuration space.
We then introduce a compressible extension by following restacking with Lagrangian relaxation to obtain the ground state, and verify the predictions against direct numerical simulations. 
Finally, we apply the method to the $m=0$ ($m$ being the poloidal mode number) interchange instability in a Z-pinch (i.e., sausage instability), which demonstrates its potential as a general framework for quantifying the nonlinear extent of macroscopic events in fusion plasmas. 
\added{In particular, its generalizability is  enhanced by formulating the restacking step as a linear sum assignment problem (LSAP)  \cite{Hieronymus2015,Stansifer2017,Harris2018,Hosking2025,hosking2026explosive} so that a local stability criterion is not necessarily required.}

\textit{Incompressible RTI}---In collisionless plasmas, the governing Vlasov equation is a continuity equation in phase space $\mathbf{z}\equiv(\mathbf{x},\mathbf{v})$. 
The flow is incompressible, meaning that the volume $\mathrm{d}\Gamma$ and the number density $f$ in a given infinitesimal phase-space element are conserved as the element moves. 
Subject to this constraint, Gardner  defines the ground state as one that minimizes the energy $W = \int f\epsilon\,\mathrm{d}\Gamma$, where the particle energy $\epsilon$ is a function of $\mathbf{z}$ only \cite{gardner1963bound}. 
It follows that the ground state $f_\text{g}$ is a monotonically decreasing function of $\epsilon$, which can be obtained by restacking the incompressible phase-space elements using standard sorting algorithms, or formally by solving the following integrodifferential equation \cite{dodin2005variational,helander2017available,helander2020available,mackenbach2022available}:
\begin{align}
\frac{\mathrm{d} f_\text{g}(\eta)}{\mathrm{d} \eta} = - \frac{\int\delta[\eta-\epsilon(\mathbf{z})]\,\mathrm{d}\Gamma}{\int\delta[f_0(\mathbf{z})- f_\text{g}]\,\mathrm{d}\Gamma},\label{eq:restackf}
\end{align}
where $f_0$ is the initial distribution function that may be unstable, $\delta(x)$ is the Dirac delta function {, and $\eta$ is a dummy variable for the particle energy $\epsilon$.} The available energy $A= W_0-W_\text{g}$ provides a nonlinear bound on the extent of the instability, where $W_0$ is the energy of $f_0$.

 {We identify an equivalent formalism in configuration space for} incompressible fluids in an external gravitational field $\phi(\mathbf{x})$, similar to the incompressible variation of Lorenz's theory \cite{Holliday1981,Winters1995}. 
The continuity equation ensures that the mass density $\rho$  {in any given} infinitesimal fluid element is conserved along with its volume $\mathrm{d}V$. 
Subject to this constraint, the ground state $\rho_\text{g}$ minimizes the energy $W = \int \rho\phi\,\mathrm{d}V$ by restacking the incompressible fluid elements, such that $\rho_\text{g}$ is a monotonically decreasing function of $\phi$ and  
\begin{align}
\frac{\mathrm{d} \rho_\text{g}(\psi)}{\mathrm{d} \psi} = - \frac{\int\delta[\psi-\phi(\mathbf{x})]\,\mathrm{d}V}{\int\delta[\rho_0(\mathbf{x})- \rho_\text{g}]\,\mathrm{d}V},\label{eq:restackrho}
\end{align}
where $\rho_0$ is the initial density distribution that may be unstable to RTI, and $\psi$ is a dummy variable for the gravitational potential $\phi$. 
The limitation of this analogy is that incompressibility is merely an approximation for fluids and plasmas. 
Next, we propose a compressible extension, which recovers the incompressible case as a limit.

\textit{Compressible RTI}---Like Lorenz \cite{Lorenz1955}, we seek a ground state that is a stable equilibrium and preserves all the local invariants of the initial state.
In compressible fluids, the continuity equation still holds. The volume and the mass density of a co-moving fluid element are not conserved, but the mass $\rho\,\mathrm{d}V$ is. 
Meanwhile, the adiabatic equation implies that the specific entropy [$s=\ln (p/\rho^\gamma)$ in an ideal gas, where $p$ is the pressure and $\gamma$ the adiabatic index] is also conserved. 
These invariants now constrain the restacking process. 
For simplicity, we consider $\phi=gx$ with a positive constant $g$ in Cartesian geometry. The initial density $\rho_0(x_0)$ and pressure $p_0(x_0)$ profiles are in force balance: $ \rho_0 g + {\mathrm{d}p_0 }/{\mathrm{d} x_0}  = 0.$

\begin{figure}
    \centering
    \includegraphics[width=1\linewidth]{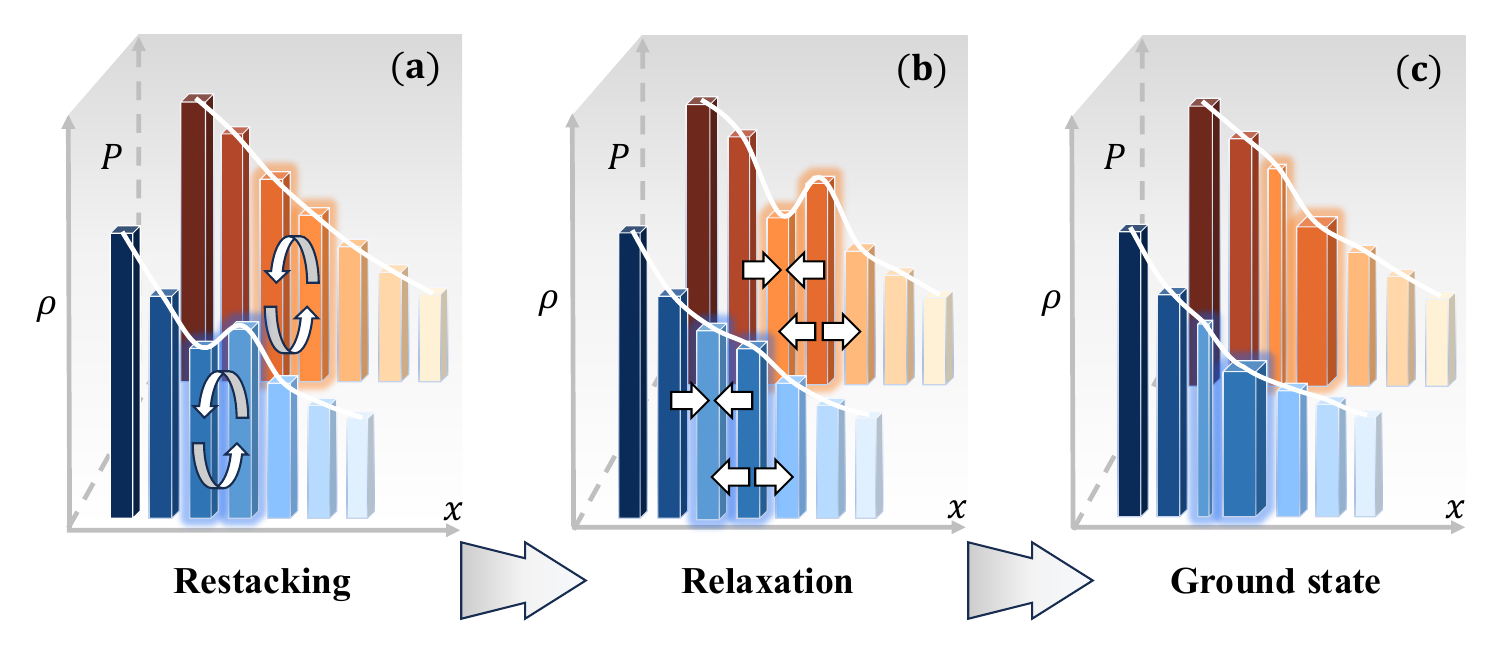}
    \caption{A schematic of the restacking--relaxation method applied to the compressible RTI: (a) the unstable initial profiles subject to incompressible restacking; (b) the restacked profiles subject to Lagrangian  relaxation; (c) the relaxed profiles, i.e., the ground state. The bars represent discrete fluid elements, whose colors and sizes denote the initial position $x_0$ and volume, respectively.}
    \label{fig:diagram}
\end{figure}

\begin{figure}[t]
    \centering

    \includegraphics[width=1\linewidth]{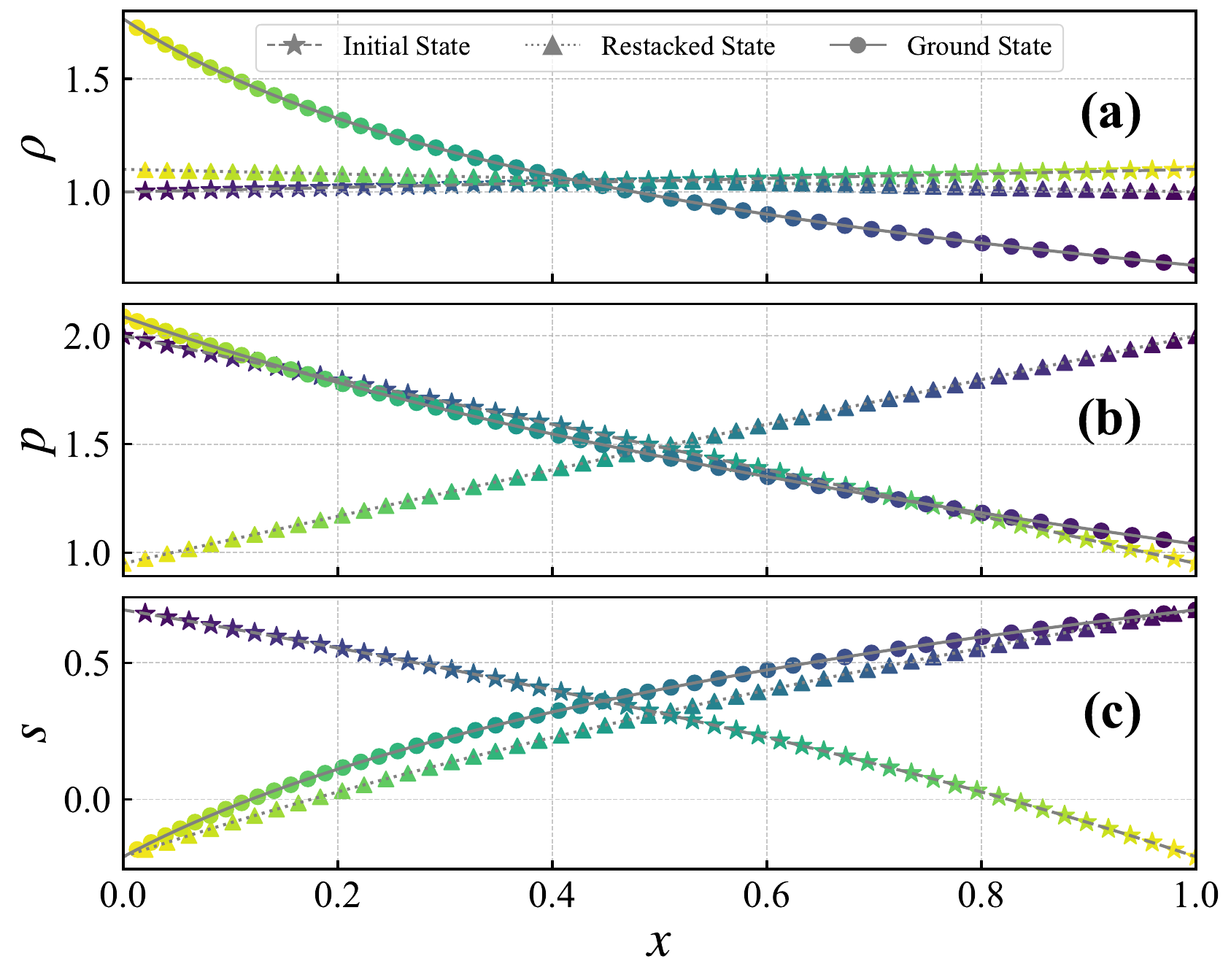} 
    
    \caption{ {Examples of initial (stars), restacked (triangles), and ground state (circles) profiles in the compressible RTI: the density (a), pressure (b), and specific entropy (c). The markers represent discrete fluid elements, whose colors denote their initial position $x_0$, like in Fig.\ \ref{fig:diagram}.}}
    \label{fig:states}
\end{figure}

\begin{figure*}
    \centering

    \includegraphics[width=\linewidth]{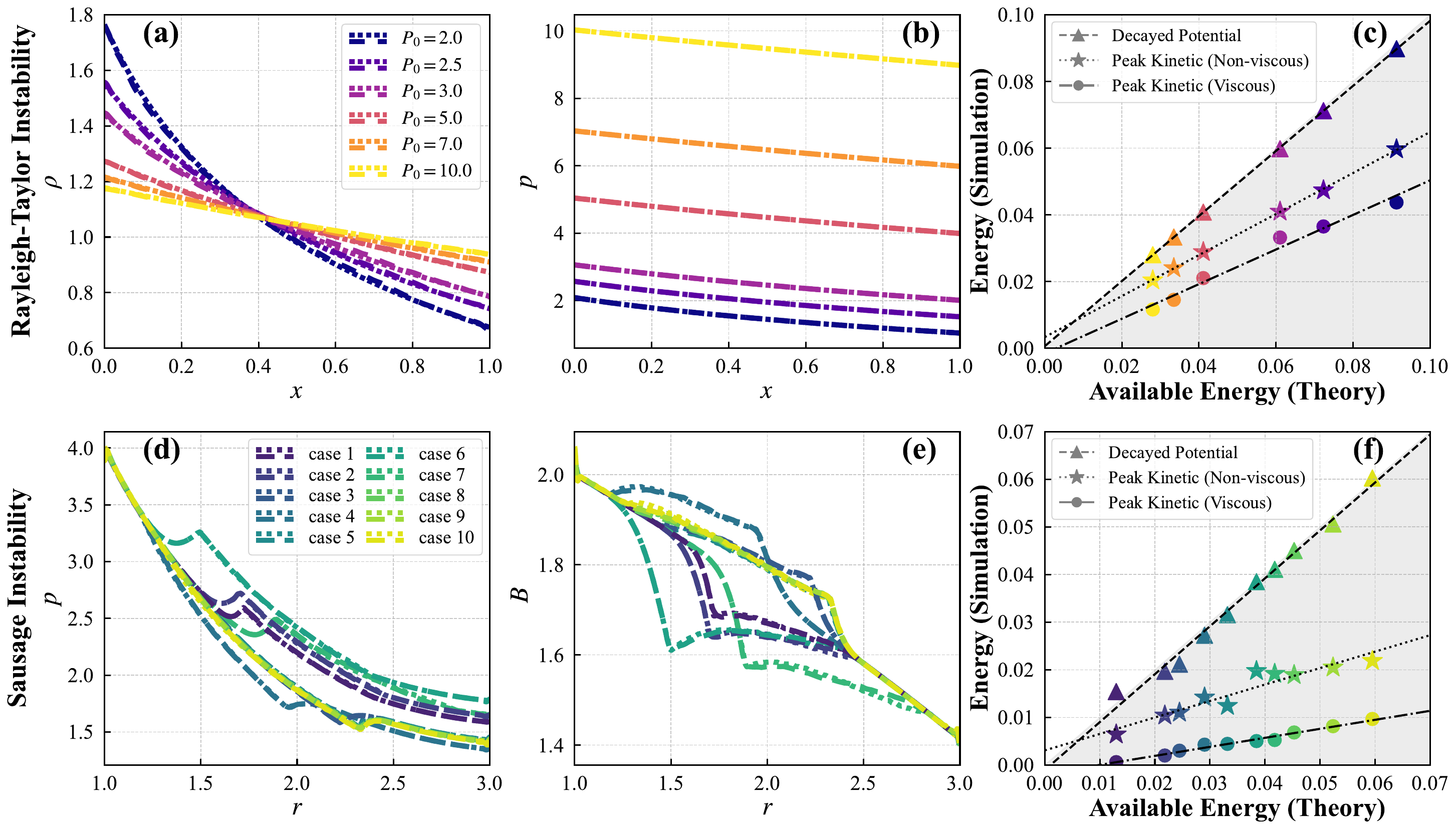}
    
    \caption{ {
    Comparison between the ground states (dotted) obtained with the proposed method and the settled profiles (dashed) in viscous simulations: density (a) and pressure (b) profiles in the RTI, and pressure (d) and magnetic field (e) in the sausage instability. The decayed potential energy in viscous simulations (triangles) and the peak kinetic energy in viscous (circles) and non-viscous simulations (stars) are all approximately proportional to the available energy for both the RTI (c) and the sausage instability (f). The black broken lines show linear fits and the shade indicates regions bounded by the available energy. 
    }} 
    \label{fig:comparison}
\end{figure*}

We find the ground state in two steps, the first being a discontinuous restacking operation as illustrated in Fig.\ \ref{fig:diagram}(a).
For convenience, we consider discrete fluid elements with the same volume so the restacking is still incompressible. 
In this case, when an element initially at $x_0$ is moved to $x_1$, its density and pressure are maintained so that $\rho_1 = \rho_0$,  $p_1 = p_0$.   
To achieve stability, we restack the fluid elements to satisfy the Schwarzschild criterion, $\mathrm{d}s/\mathrm{d}x>0$, a necessary and sufficient condition for linear stability \cite{KULL1991197}. 
Since $s$ is invariant, similar to Eqs.\ \eqref{eq:restackf} and \eqref{eq:restackrho}, the restacked state is formally given by
\begin{align}
\frac{\mathrm{d} s_1(x_1)}{\mathrm{d} x_1} =  \frac{1}{\int\delta[s_0(x_0)- s_1]\,\mathrm{d}x_0},\label{eq:restacks}
\end{align}
where $s_0(x_0)$ and $s_1(x_1)$ denote the initial and restacked specific entropy profiles, respectively. 
In Fig.\ \ref{fig:states}, we show an example of the restacked profiles $\rho_1(x_1)$, $p_1(x_1)$, and $s_1(x_1)$ (triangles) to compare with the unstable initial profiles (stars),  {where the fluid elements are color-coded based on their initial locations ($x_0$) to track their spatial rearrangements.}

Unlike the incompressible case, the restacked state produced at this first step is not in force balance. 
To further obtain an equilibrium state with lower energy, we adopt Lagrangian relaxation as illustrated in Fig.\ \ref{fig:diagram}(b). 
Specifically, the fluid motion is governed by a continuous mapping $x_\text{g}(x_1)$, and the invariance of mass and entropy implies that \cite{Newcomb1962}
\begin{equation}
     \rho_\text{g}[x_\text{g}(x_1)]={\rho}_1/{J},\quad  p_\text{g}[x_\text{g}(x_1)] =  {p}_1/{J^\gamma},\label{eq:advection}
\end{equation}
where the Jacobian     $J(x_1) = {\mathrm{d} x_\text{g}}/{\mathrm{d} x_1}$ captures the compressibility of the fluid.
Minimizing the potential energy $W = \int [\rho_\text{g} x_\text{g}g+p_\text{g}/(\gamma-1)]\,\mathrm{d}x_\text{g}$ yields the relaxed state as portrayed in Fig.\ \ref{fig:diagram}(c), which satisfies force balance:
\begin{equation}
    \rho_1(x_1)g
    + \frac{\mathrm{d}}{\mathrm{d} x_1}
    \left[ \frac{p_1(x_1)}{J^{\gamma}} \right] = 0. \label{eq:rtieq}
\end{equation}
The equilibrium is stable because the monotonicity of the specific entropy is preserved, $\mathrm{d}s_\text{g}/\mathrm{d}x_\text{g}=J^{-1}\mathrm{d}s_1/\mathrm{d}x_1>0$.
We regard this invariant-preserving stable equilibrium as the ground state, based on which the available energy can be calculated.
The corresponding profiles $\rho_\text{g}(x_\text{g})$ and $p_\text{g}(x_\text{g})$ in Fig.\ \ref{fig:states} (circles) show significant deviation from the restacked state as needed to reach force balance.

To verify the method's predictions, we perform direct numerical simulations using the \texttt{Dedalus} framework \cite{burns2020dedalus}. 
We solve the compressible fluid equations in 2D slab geometry $(x,y)$.
The boundary conditions are Neumann for the pressure and density and free-slip for the velocity in $x$, and periodic in $y$. 
Small hyper-diffusivities are included in the continuity and adiabatic equations for numerical stability. A large viscosity is used to damp out the RTI-driven flows so that the system settles to a near-equilibrium state. 
The initial conditions are based on the globally unstable profiles in Fig.\ \ref{fig:states} (stars),  {$\rho_0 = 1+0.1x_0$ and $p_0 = P_0 -x_0 - 0.05x_0^2$}, with the maximum pressure $P_0$ varied to adjust compressibility. 
 {Further details of the simulation settings are provided in the End Matter}.

In Fig.\ \ref{fig:comparison}(a-b), all the settled profiles (dashed) show excellent agreement with the corresponding ground-state profiles (dotted) obtained with the proposed method, corroborating that the latter are consistent with the constrained energy-minimizing equilibria. 
This naturally translates to the agreement between the decayed potential energy (triangles) in the simulations and the calculated available energy shown in Fig.\ \ref{fig:comparison}(c). 
In particular, the most compressible case ($P_0=2$) corresponds to Fig.\ \ref{fig:states}, and the least compressible case ($P_0=10$) approaches the incompressible limit. 
 {Additionally, a parameter scan is shown in the End Matter to demonstrate that the agreement holds for a broad range of viscosity.}

 {To verify the available energy's effectiveness as a metric in a dynamic setting, we measure the peak kinetic energy in these ``viscous" simulations as well as another series of ``non-viscous'' simulations with zero viscosity and a small hyper-viscosity. 
In Fig.\ \ref{fig:comparison}(c), the peak kinetic energy is shown to be roughly proportional to the available energy in both scenarios. 
This supports using the available energy to estimate the nonlinear extent of instabilities, in particular the magnitude of transient events induced, such as ELMs or sawteeth. The latter tend to be a major concern for macroscopic instabilities in fusion plasmas.}

\textit{Sausage instability}---Using the restacking--relaxation method, one can obtain the ground state robustly and efficiently in more general cases as well, irrespective of the specific geometry or types of local invariants.
We use the sausage instability as an example to show how to treat magnetized fluids and more complex geometry. 
It is an axisymmetric ($m=0$) interchange mode in a Z-pinch, which features a purely poloidal magnetic field $\mathbf{B} = B\mathbf{e}_{\theta}$ in cylindrical geometry $(r,\theta,z)$. 
 {For simplicity, here we still focus on 1D equilibrium profiles $p_0(r_0)$ and $B_0(r_0)$, even though the method can in general treat initial states that are multidimensional or not in force balance.}

In ideal magnetohydrodynamics (MHD), the magnetic flux $\mathbf{B}\cdot \mathrm{d}\mathbf{S}$ through an area element $\mathrm{d}\mathbf{S}$  co-moving with the plasma is conserved \cite{Newcomb1962}. 
This entails that $B_1\,\mathrm{d}r_1=B_0\,\mathrm{d}r_0$ when we radially rearrange a fluid element from $r_0$ to $r_1$. 
Similar to the RTI case, we resort to the linear stability  criterion \cite{kadomtsev1966hydromagnetic} to formulate a restacking strategy.
The necessary and sufficient condition can be written in a form like the Schwarzschild criterion  {(cf.\ End Matter)}: 
\begin{equation}
\frac{\mathrm{d}\sigma}{\mathrm{d} r} > 0, \qquad \sigma =\ln\! \left( \frac{rp^{1/\gamma}}{B} \right).
   \label{eq:sausagef}
\end{equation}
Note that $\sigma$ is also conserved in 1D dynamics due to the conservation of mass, entropy, and magnetic flux, so the elements can be sorted directly in ascending order of $\sigma$. 

\begin{figure}[t]
    \centering

    \includegraphics[width=1\linewidth]{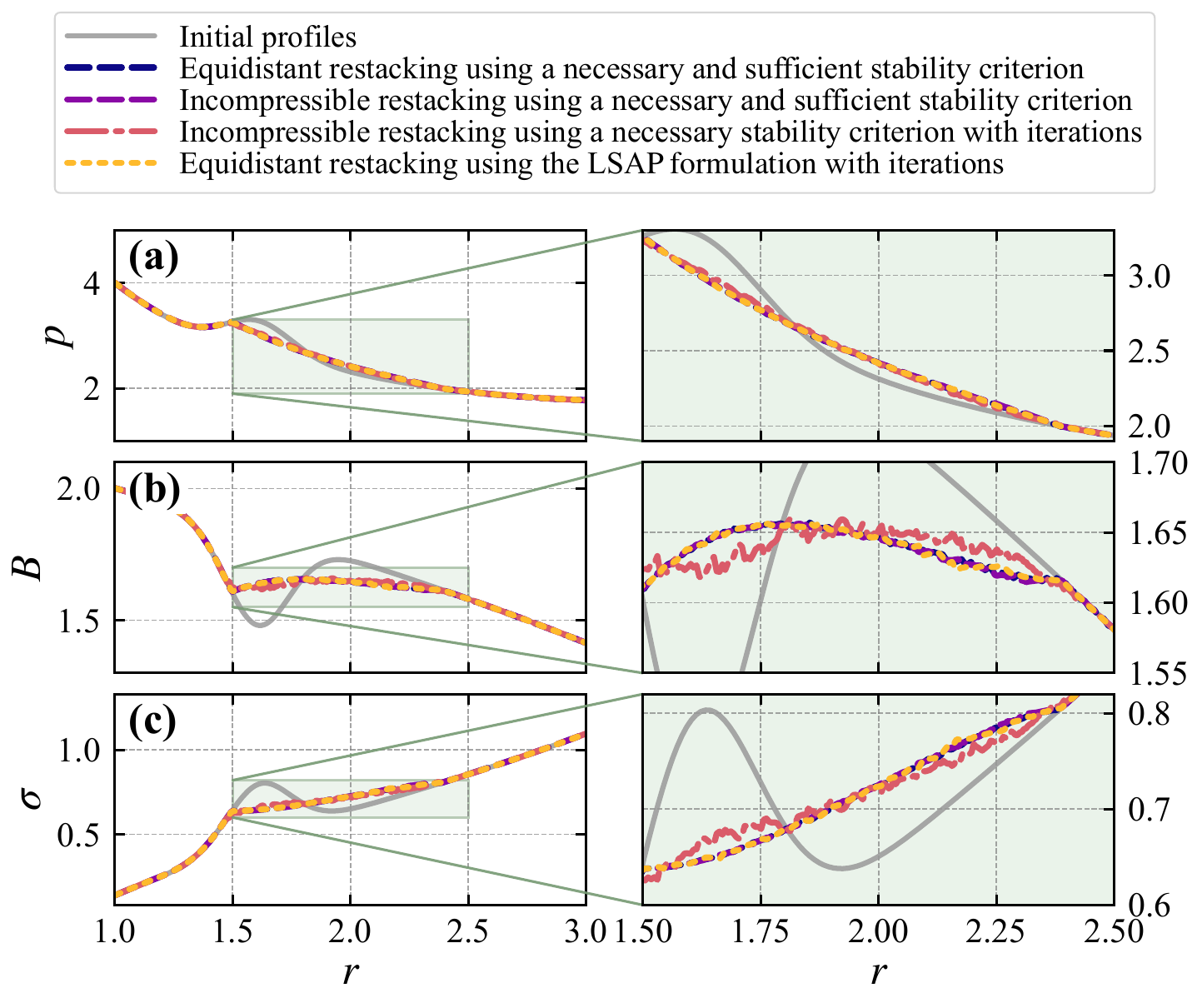} 
    
    \caption{ {The final relaxed-state pressure (a), magnetic field (b), and stability indicator (c) profiles (broken) obtained using different restacking strategies show good agreement. The gray solid curves show the initial profiles. The right column shows enlargements of the boxed areas in the left column.}}
    \label{fig:3 schemes}
\end{figure}

Here, we can also consider discrete elements with the same volume $\mathrm{d}V {\propto}r\mathrm{d}r$, so the restacking operation is still incompressible. In this case, we have $B_{1}/r_1 = B_0/r_0$ and $p_1=p_0$, and the restacked state is formally given by
\begin{equation}
    \frac{\mathrm{d} \sigma_1}{\mathrm{d} r_1} = \frac{r_1}{\int \delta[\sigma_0(r_0)-\sigma_1]r_0 \mathrm{d}r_0}.\label{eq:restacksigma1}
\end{equation}
 {However, different discretizations are allowed if needed.
For example,} we can consider discrete elements with the same radial extent $\mathrm{d}r$, so that the restacking operation is ``equidistant". In this case, we have $p_1 / r_1^\gamma = p_0 / r_0^\gamma$ and $B_{1} = B_0$, and the restacked state formally satisfies
\begin{equation}
    \frac{\mathrm{d} \sigma_1}{\mathrm{d} r_1} = \frac{1}{\int \delta[\sigma_0(r_0)-\sigma_1] \mathrm{d}r_0}.\label{eq:restacksigma2}
\end{equation}
The restacked states in the two cases will have different profiles but the same order of fluid elements by ascending $\sigma$, so the ground states then obtained by Lagrangian relaxation are the same, as illustrated in Fig.\ \ref{fig:3 schemes}. 
Similar to Eq.\ \eqref{eq:rtieq}, the equilibrium that minimizes the potential energy $W = \int [B^2_\text{g} /2+p_\text{g}/(\gamma-1)]\,r_\text{g}\mathrm{d}r_\text{g}$ reads
\begin{equation}
    \frac{\mathrm{d}}{\mathrm{d} r_1} \left(  \frac{p_1}{J^{\gamma}}  + \frac{r_{\text{g}}^2}{r_1^2 } \frac{B_1^2}{2J} \right) + \frac{B_1^2}{r_1 J}=0,
\end{equation}
where $J=({r_{\text{g}}}/{r_1})({\mathrm{d} r_{\text{g}}}/{\mathrm{d} r_1})$. The solution $r_\text{g}(r_1)$ then yields the ground state profiles
\begin{equation}
    B_{\text{g}}[r_\text{g}(r_1)]= r_{\text{g}} B_1/r_1 J,\qquad p_{\text{g}}[r_\text{g}(r_1)]=p_1/J^\gamma.
\end{equation}

The RTI and the sausage instability each admit an explicit local necessary and sufficient stability criterion that can be expressed in terms of a local invariant. 
This feature simplifies the restacking process but may not exist in general. 
However, the restacking strategy does not require a necessary and sufficient criterion--in Gardner's restacking algorithm, ${\mathrm{d} f_\text{g}}/{\mathrm{d} \epsilon}<0$ is only a sufficient condition for stability. (The necessary and sufficient condition was given by Penrose \cite{Penrose1960}.) 
In Fig.\ \ref{fig:3 schemes}, we show the result obtained using a necessary stability criterion, $\mathrm{d}(B^2/r^2)/\mathrm{d}r<0$, which is close to the ground state obtained using the necessary and sufficient criterion. 
Here, the restacking--relaxation method is applied iteratively until convergence (cf.\ Fig.\ \ref{fig:iteration} in End Matter), since the criterion is not given in terms of a local invariant and hence the relaxed state can be unstable after one iteration. 
In addition, Fig.\ \ref{fig:3 schemes} also shows the ground state obtained by formulating the restacking step as an LSAP \cite{Hieronymus2015,Stansifer2017,Harris2018,Hosking2025,hosking2026explosive} and directly minimizing the energy over parcel assignments without invoking a stability criterion (cf.\ End Matter). 
In this case, the ground state is achieved after only a few iterations and agrees almost exactly with those obtained using the necessary and sufficient criterion.
\added{The LSAP-based restacking strategy significantly enhances the flexibility and generalizability of the proposed framework by removing the requirement for a specific form or the existence of a local stability criterion.}

To verify the predictions, we solve the ideal MHD equations in axisymmetric cylindrical geometry $(r,z)$ using \texttt{Dedalus}. 
The simulation settings are similar to those for the RTI  {(see End Matter for details). 
The initial profiles are based on the example in Fig.\ \ref{fig:3 schemes} (case 6) with the position and shape of the pressure hump (or hollow) varied.  
In Fig.\ \ref{fig:comparison}(d-e), all the settled profiles (dashed) in the simulations agree well with the ground states (dotted) predicted.
In Fig.\ \ref{fig:comparison}(f), the decayed potential energy in viscous simulations and the peak kinetic energy in viscous and non-viscous simulations all show approximate proportionality to the available energy.} 
The agreement is less perfect than the globally unstable RTI cases in Fig.\ \ref{fig:comparison}(c) because here the available energy takes up a much smaller fraction of the total energy, so that the results are more sensitive to numerical settings,  {the boundary conditions in particular}.
Nevertheless, the clearly observable trend supports the use of the available energy to estimate the nonlinear extent of the instability.

\textit{Summary}---In this work, we present a general method for calculating the available energy as a metric for the nonlinear extent of convective hydrodynamic and hydromagnetic instabilities. 
The method extends Gardner's restacking algorithm to configuration space and combines it with Lagrangian relaxation to treat compressibility. Its effectiveness and generalizability are demonstrated by successful application to the RTI and the sausage instability. 
Future work should consider more complicated instabilities in realistic geometries, such as interchange modes in a screw pinch and ultimately ballooning modes in toroidal plasmas, \added{for which a multidimensional treatment is likely needed and the LSAP-based restacking strategy might be more useful}.

\textit{Acknowledgments}---The authors are grateful for useful discussions with Y. Chen, Q. Sun, H. Zhu, X. Xu, X. Wu,  G. Jia, D. N. Hosking, S. de Pirey, and I. Y. Dodin. This research is supported by the National MCF R\&D Program under Grant Number 2024YFE03230400, the National Natural Science Foundation of China under Grant Number 12305246, and the Fundamental Research Funds for the Central Universities.

\bibliography{my} 


\clearpage
\section*{End Matter}

\appendix

\textit{Limitation of Lorenz's method---}For a stratified atmosphere, Lorenz \cite{Lorenz1955} considered a hydrostatic reference or ground state, ${\mathrm{d}p}/{\mathrm{d}x}=-\rho g $. With $p(\infty)=0$, the force-balance equation reduces to a simple algebraic relation
\begin{equation}
    p(x)   =g\int_x^{\infty}\rho(x')\,\mathrm{d}x'= M(x) g,\label{eq:masscoord}
\end{equation}
which introduces the mass coordinate $M$ that is central to Lorenz's method. With $\mathrm{d}M = -\rho \mathrm{d}x $, the total potential energy can then be expressed as \cite{Hosking2025}
\begin{subequations}
\begin{align}
    W &=\int_0^{\infty}
    \left[
        \frac{p}{(\gamma-1)}
        +\rho gx
    \right]\mathrm{d}x\\
    &=\int_0^{M_{\mathrm{tot}}}
    \left[
        \epsilon(p,s)
        +\frac{Mg}{\rho}
    \right]\mathrm{d}M\label{eq:pot_in_mass}\\
    &=\int_0^{M_{\mathrm{tot}}}
        h(Mg,s)\,\mathrm{d}M \label{eq:entalpy},
\end{align}
\end{subequations}
where $M_{\mathrm{tot}}$ is the total mass, and $\epsilon$ and $h$ are the specific internal energy and specific enthalpy, respectively, both expressed in terms of $(p,s)$.
Here, integration by parts is used to get Eq.\ \eqref{eq:pot_in_mass}, and Eq.\ \eqref{eq:masscoord} is used to get Eq.\ \eqref{eq:entalpy}. 
Then, minimizing $W$ can be formulated as adiabatically sorting fluid parcels in the mass coordinate $M$. \added{Subsequent work extended Lorenz’s formulation to include moisture and magnetic field, and recognized the parcel-sorting problem as an LSAP that can be solved by established algorithms \cite{Lorenz1979,Hieronymus2015,Stansifer2017,Harris2018,Hosking2025}}. 

In many other cases, the force balance condition is more complicated and cannot be reduced to a simple algebraic relation like Eq.\ \eqref{eq:masscoord}, but instead remains in differential form. Therefore, we propose treating compressibility using Lagrangian relaxation, which can handle force balance equations in differential form.

\textit{Simulation settings}---In the RTI simulations, we solve the dimensionless compressible hydrodynamic equations in 2D slab geometry $(x,y)$ with $\mathbf{v}=v_x\,\mathbf{e}_x + v_y\,\mathbf{e}_y$:
\begin{subequations}
\begin{align}
    &{\partial_t \rho}+ \nabla\cdot(\rho \mathbf{v}) = -{D_\text{h}} \nabla^4 \left(\rho-\rho_0\right),\\
    &{\partial_t p}+ \mathbf{v}\cdot\nabla p + \gamma p \nabla\cdot \mathbf{v}= -\kappa_\text{h}\nabla^4 \left(p-p_0\right),\\
    &{\rho} \left({\partial_t \mathbf{v}}+ \mathbf{v} \cdot \nabla \mathbf{v} \right)  = -\nabla p  - \rho{g}\mathbf{e}_{x} + \nu \nabla^2 \mathbf{v} - \nu_\text{h}\nabla^4\mathbf{v}, 
\end{align}
\end{subequations}
where $\gamma=5/3$, ${g}=1$, {$\nu$} is viscosity and $D_\text{h},\, \kappa_\text{h}, \, \nu_\text{h}$ are hyper-diffusivities adopted for numerical stability.  In viscous simulations, $D_\text{h}=5.0\times10^{-12},\, \kappa_\text{h}=10^{-11}, \, \nu=3.0\times10^{-3}$, and $\nu_\text{h}=0$. In non-viscous simulations, $D_\text{h}=5.0\times10^{-11},\, \kappa_\text{h}=2\times10^{-11}, \, \nu=0,$ and $ \nu_\text{h}=5.0\times10^{-11}$. 
\texttt{Dedalus} uses the pseudospectral method and we choose Chebyshev and Fourier bases in $x$ and $y$, respectively. The simulation domain is $[0,1]^2$ with the resolution $(N_x,N_y)$ being $(260,80)$ and $(600,100)$ for the viscous and non-viscous simulations, respectively.

For the sausage instability, we solve the dimensionless MHD equations in axisymmetric cylindrical geometry $(r,\,z)$ with $\mathbf{B}=B\,\mathbf{e}_\theta$ and $\mathbf{v}=v_r\,\mathbf{e}_r + v_z\,\mathbf{e}_z$:
\begin{subequations}
\begin{align}
& {\partial_t \rho}+ \nabla\cdot(\rho \mathbf{v}) = -{D_\text{h}}\nabla^4 \left(\rho-\rho_0\right),\\
& {\partial_t p}+ \mathbf{v}\cdot\nabla p + \gamma p \nabla\cdot \mathbf{v}= -\kappa_\text{h}\nabla^4 \left(p-p_0\right), \\
&    \partial_t \mathbf{B}  = {\nabla}\times ({\mathbf{v}} \times {\mathbf{B}}  ) - \eta_\text{h} \nabla^4 ({\mathbf{B}}  - \mathbf{B}_0),\\
&    \rho \left({\partial_t \mathbf{v}}+ \mathbf{v} \cdot \nabla \mathbf{v} \right)   = -\nabla p + (\nabla\times\mathbf{B})\times \mathbf{B}\nonumber
\\&  \qquad \qquad \qquad \qquad \ \     +\nu\nabla^2 \mathbf{v}- \nu_\text{h}\nabla^4\mathbf{v}, 
\end{align}
\end{subequations}
where $\eta_\text{h}$ is a hyper-resistivity. In viscous simulations, $D_\text{h}=10^{-11},\  \kappa_\text{h}=10^{-12},\ \eta_\text{h} = 10^{-11},\ \nu=5\times 10^{-3},$ and $ \nu_\text{h} = 0$. In non-viscous simulations, $D_\text{h} = 6.0\times 10^{-11},\ \kappa_\text{h}=5.0\times 10^{-11},\ \eta_\text{h}=6.0\times 10^{-11},\, \nu = 0$, and $ \nu_\text{h}=3.0\times 10^{-11}$.
We choose Chebyshev basis in $r$ with Neumann boundary conditions for $\rho,\,p,\,rB$ and free-slip for $\mathbf{v}$, and Fourier basis with periodic boundary conditions in $z$.
The simulation domain is $[1,3]\times[0,1]$ with the resolution $(N_r,N_z)$ being $(400,80)$ and $(640,80)$ for the viscous and non-viscous simulations, respectively. 
The initial profiles are
\begin{subequations}
\begin{align}
    B_0^2 &= 5-r_0-A_B \exp\left[-\frac{(r_0-r_B)^2}{\Delta r^2}\right], \\
    p_0 &= 5-\frac{B_0^2}{2}-5\ln(r_0)+r_0 \notag\\
    &\quad +\int_{1}^{r_0}\frac{A_B}{r'} \exp\left[-\frac{(r'-r_B)^2}{\Delta r^2}\right]\,\mathrm{d}r',
\end{align}
\end{subequations}
and $\rho_0 = p_0$. The parameters $A_B,\,r_B,\,\Delta r$ are varied to generate the cases shown in Fig.\ 3(d,e).

In all cases, the simulations are initialized with a small 2D Gaussian velocity perturbation. The profiles shown in Fig.\ \ref{fig:comparison} are all averaged in the periodic directions ($y$ and $z$, respectively).

\begin{figure}
    \centering
    \includegraphics[width=1.0\linewidth]{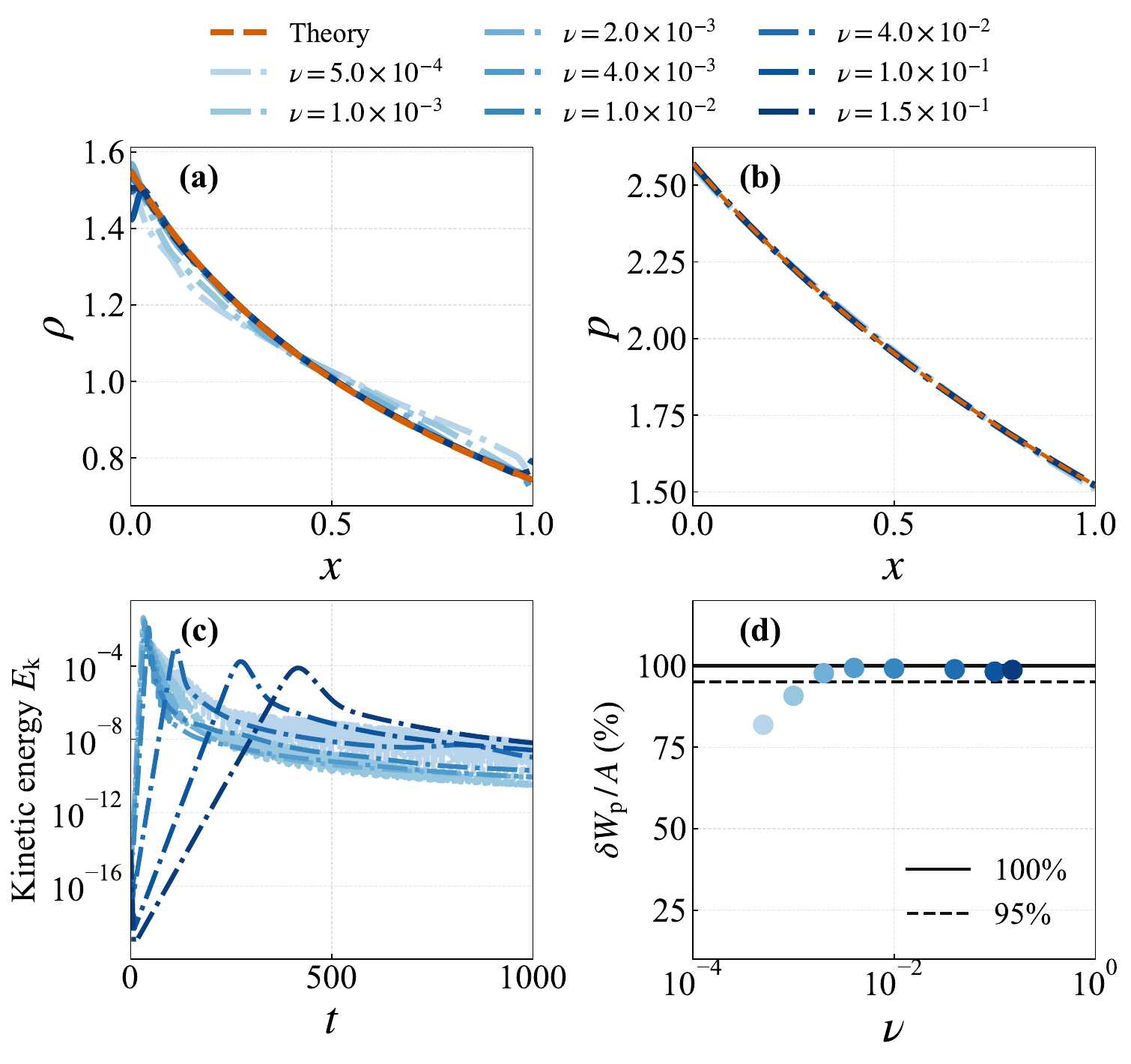}
    \caption{The saturated density (a) and pressure (b) profiles (dash-dot) in RTI simulations with varying viscosity $\nu$, compared with the theoretical ground state (dashed). Panel (c) shows the time evolution of the kinetic energy $E_\text{k}$. Panel (d) shows the fraction of decayed potential energy $\delta W_\text{p}$ over the available energy $A$ versus $\nu$, demonstrating convergence to the theoretical limit over a broad range of viscosity.}  
    \label{fig:viscosity scan}
\end{figure}

\textit{Viscosity scan}---To illustrate the influence of viscosity $\nu$ on the viscous RTI simulation results, we perform a parameter scan as shown in Fig.\ \ref{fig:viscosity scan}. Panels (a) and (b) present the final density and pressure profiles in simulations with $\nu$ varying from $5.0\times 10^{-4}$ to $1.5\times 10^{-1}$ for the $P_0=2.5$ case in Fig.\ \ref{fig:comparison}, compared with the ground state predicted by the restacking--relaxation method (dashed). Panels (c) and (d) display the time evolution of the kinetic energy and the ratio of the decayed potential energy to the available energy, respectively. 

In these simulations, the primary role of viscosity is to dissipate the system's kinetic energy to reach a nearly static state. Therefore, a relatively large $\nu$ is needed for efficient damping within practical simulation time.
As Fig.\ \ref{fig:viscosity scan} shows, over a broad range of viscosity, $2\times10^{-3}\le \nu \le 1.5\times10^{-1}$, the relaxed state is remarkably robust and the decayed potential energy consistently converges to the theoretical prediction. 
This demonstrates that the verification of the theoretical ground states by direct numerical simulations presented in Fig.\ \ref{fig:comparison} is not sensitive to the specific value of viscosity.

\textit{Criterion for sausage instability}---The necessary and sufficient linear stability criterion given in Ref.\,\cite{kadomtsev1966hydromagnetic} is 
\begin{equation}
    \frac{2\gamma B^2}{\gamma p+B^2 } + \frac{r}{p}\frac{\mathrm{d} p}{\mathrm{d} r} > 0,
   \label{eq:sausage}
\end{equation}
which can be rewritten as 
\begin{equation}
    \gamma \left( \frac{\mathrm{d} p}{\mathrm{d}r} + \frac{2 B^2}{r}  \right)  + \frac{\mathrm{d} p}{\mathrm{d} r} \frac{B^2}{p}>0. 
\end{equation}
Using the force balance condition, 
\begin{equation}
    \frac{\mathrm{d}}{\mathrm{d} r} \left(p + \frac{B^2}{2}\right)
    + \frac{B^2}{r} = 0,\label{eq:force_balance}
\end{equation}
we can obtain
\begin{equation}
    \gamma \left(\frac{ B^2}{r} - B \frac{\mathrm{d}B}{dr}  \right)  + \frac{\mathrm{d} p}{\mathrm{d} r} \frac{B^2}{p}>0, 
\end{equation}
or equivalently,
\begin{equation}
    {\frac{\mathrm{d} p}{\mathrm{d} r}  \frac{2}{\gamma p}+\frac{B^2}{r^2}\frac{\mathrm{d}}{\mathrm{d}r}\left( \frac{r^2}{B^2}\right)>0,}
\end{equation}
which yields Eq.\ \eqref{eq:sausagef}, that is, $\mathrm{d} \sigma/{\mathrm{d} r}>0$. The conservation of $\sigma$ can be shown by combining the conservation of mass ($\rho r\mathrm{d}r$), entropy ($p/\rho^\gamma$), and magnetic flux ($B\mathrm{d}r$).

\begin{figure}
    \centering
    \includegraphics[width=1\linewidth]{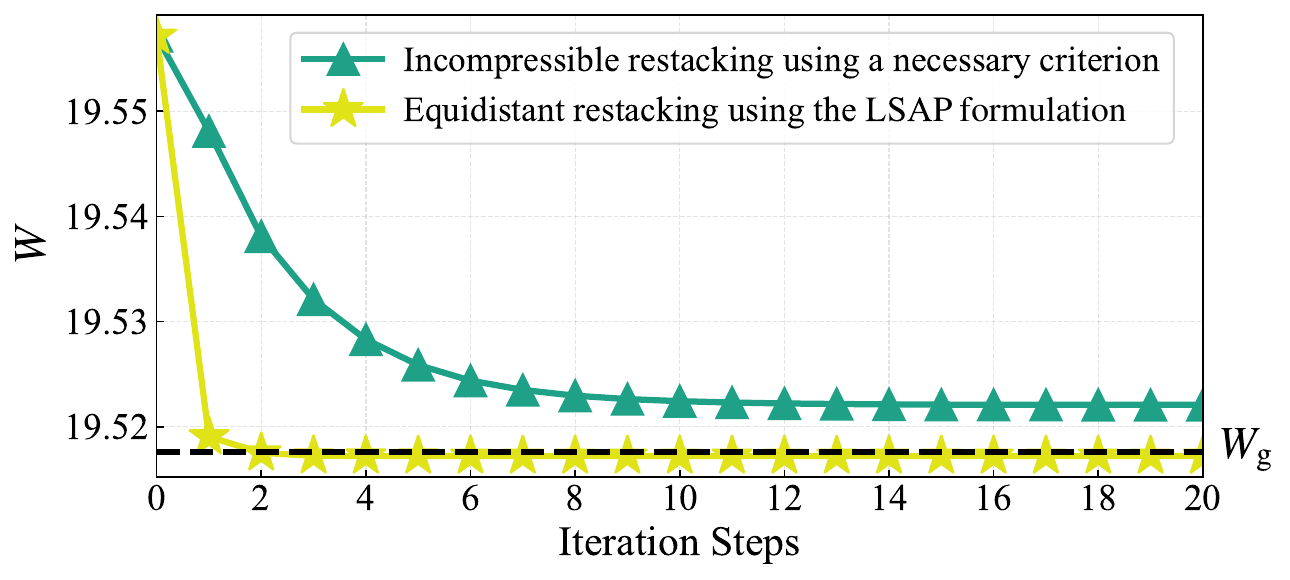}
    \caption{The evolution of the total potential energy $W$ when the proposed method is applied iteratively, using a necessary criterion (triangles) or the LSAP formulation (stars) for restacking. The black dashed line marks the ground state value $W_\text{g}$.}
    \label{fig:iteration}
\end{figure}

\textit{Convergence by iterations}---In Fig.\ \ref{fig:3 schemes}, the final relaxed states obtained with different restacking strategies are shown, including two where the restacking--relaxation method is applied iteratively.
In Fig.\ \ref{fig:iteration}, we show how the total potential energy $W$ changes with iterations. For incompressible restacking with a necessary criterion, $W$ gradually converges to a value that is close to but different from the ground state value $W_\text{g}$. 
For equidistant restacking with the LSAP formulation, $W$ converges to $W_\text{g}$ almost exactly and more quickly.
One can see that iterations are necessary and effective in these cases.

\textit{LSAP-based restacking}---Following Refs.\,\cite{Hosking2025, hosking2026explosive}, one can also formulate the restacking step
as an LSAP in the sausage instability. \added{This strategy does not require a stability criterion and hence enhances the generalizability of the proposed framework.} 
When moving a discrete fluid element from its original location $r_j$ to a target location $r_i$, the corresponding element of the cost matrix $C_{ij}$ is defined by its potential energy after moving,
\begin{equation}
  {  C_{ij} = \left( \frac{p_{ij}}{\gamma -1} + \frac{B_{ij}^2}{2} \right) r_i \mathrm{d}  r_i,}
\end{equation}
where $\mathrm{d} r_i$ denotes the radial length of the element at $r_i$.
The invariance of entropy and magnetic flux implies that
\begin{equation}
   {p_{ij} = p_j \left(\frac{r_j \mathrm{d} r_j}{r_i \mathrm{d} r_i} \right)^\gamma, \quad B_{ij} = B_j \frac{\mathrm{d} r_j}{\mathrm{d} r_i} .}
   \label{eq: p and B in LSA}
\end{equation}
For equidistant restacking, $\mathrm{d} r_i = \mathrm{d} r_j$, Eq.\ \eqref{eq: p and B in LSA} becomes
\begin{equation}
   {p_{ij} = p_j \left(\frac{r_j }{r_i } \right)^\gamma, \quad B_{ij} = B_j .}
\end{equation}
For incompressible restacking, $r_i\mathrm{d} r_i = r_j \mathrm{d} r_j $, Eq.\ \eqref{eq: p and B in LSA} reduces to
\begin{equation}
    p_{ij} = p_j,\quad B_{ij} = B_j \frac{r_i}{r_j}.
\end{equation}
The classic Hungarian algorithm \cite{Kuhn1955,Munkres1957} is then used to solve this assignment problem of minimizing the total potential energy, yielding a restacked state. Lagrangian relaxation is then applied to obtain the ground state.

\end{document}